# Arbitrary mixture of two charged interacting particles in a magnetic Aharonov-Bohm ring: persistent currents and Berry's phases

**K Kyriakou[1], K Moulopoulos[1], A V Ghazaryan[2*] and A P Djotyan[2]**
[1]Department of Physics, University of Cyprus, P.O. Box 20537, 1678 Nicosia, Cyprus
[2]Department of Physics, Yerevan State University, 1 Al. Manoogian, Yerevan, Armenia

E-mail: a.ghazaryan@ysu.am, ghazaryan_areg@yahoo.com

**Abstract**. Aharonov-Bohm Physics at the two-particle level is investigated for distinguishable interacting charged particles through the exact solution of a toy model with confined states. The effect of the inaccessible magnetic flux is distributed between the center-of-mass and the internal pair level, and the nontrivial manner in which the two levels are mutually affecting each other demonstrates the interplay between interactions, nontrivial topology, the Aharonov-Bohm flux and the characteristics of a charged quantal mixture. Analytical expressions for energy spectra, wavefunctions, (flux-dependent) critical interactions for binding and current densities are derived, and these offer the rare possibility to study persistent currents from the point of view of an interacting nanoscopic system. Two cyclic adiabatic processes are identified, one coupled to the center-of-mass behavior and the other defined on the two-body interaction potential, with the associated Berry's phases also analytically determined; these are found to be directly linked to the electric and probability (persistent) currents in nontrivial ways that are shown to be universal (independent of the actual form of the interaction). The direct connection of the two-body Berry's phase to the electric current for a neutral system is found to disappear in case of identical particles – hence revealing the character of a charged mixture as being crucial for exhibiting these universal behaviors.

## 1. Introduction

The Aharonov-Bohm (AB) effect [1] (both the well-known magnetic version but also its electric counterpart) is a one-body phenomenon, in the sense that it results from the influence of (vector and scalar) potentials on the quantum mechanical phases of wavefunctions of a system with a specific charge (that actually appears in the phase); the system is either a single particle (the usual case), or in the many-particle context, the total charge may be used, in which case the AB effect concerns only the center-of-mass (CM) wavefunction. The present investigation is devoted to a more careful look of how the potentials are affecting further details of the many-particle wavefunction. As a first step we deal with the magnetic AB version and with only two particles; our model system is then exactly solvable and this gives us the luxury of studying the details of the AB influence on the complete two-particle wavefunction. This system gives us also the opportunity to study exactly the interplay between nontrivial topology and interactions, in combination with the properties of a charged mixture, and leads to results that, at the level of wavefunctions, could be used in effective two-particle descriptions of more complex many-body systems under analogous magnetic AB conditions.

---

[*] The Corresponding author

Our model system consists of two (spinless) particles (with arbitrary masses and charges) interacting through a contact potential and moving in a one-dimensional (1D) ring threaded by a (static) magnetic flux. We first determine the precise manner in which the AB effect is experienced at the CM and the internal pair level, and by way of an example (of how the two levels affect each other through the AB flux) we provide a condition on masses and charges for which the AB influence disappears from the internal physics of the pair but still operates at the CM level. We then proceed with the full solution and determine exactly the complete AB influence on energy spectrum and other properties (all determined analytically).

On the experimental side, one might mention works referring to "two-particle AB effect", although these have always concerned identical particles, and essentially non-interacting ones (by placing importance on the issue of entanglement between them [2, 3]). Technologically speaking it is recent advances in the fabrication of semiconductor nanostructures that really make the construction of III-V semiconductor volcano shaped nanorings possible [4, 5]. A brief review on types and methods of fabrication, and fascinating physical properties of nano-scale semiconductor rings is given in [6]. These nanoscopic rings may be the best suited quantum structures for investigating the electronic and optical properties of quantum rings, because they are in the scattering-free and few-particle limit [7].

The experimental observation of AB oscillations and persistent currents first predicted in Ref. [8] in small conducting rings [9, 10], on one hand, and the recent experimental developments in manufacturing quantum dots [11] and rings [12] with only a few electrons, on the other, have made quantum rings an ever increasing topic of experimental research and a new playground for many-particle theory in quasi-1D systems. Concerning non-identical particles, the system of an exciton inside a nanoring in the presence of AB fluxes was considered with various methods in [13-17] for different types of interaction potentials.

Let us focus in particular on [17], where the exciton system in a nanoring (for spinless oppositely-charged particles with equal masses) was studied theoretically for contact potential interaction with the use of a Green's function (GF) procedure. The present study extends and generalizes the results of the above work to the case of two distinct particles with arbitrary masses and charges. The new results are either specific to the system, or they present universal patterns that at the end are found to be independent of the form of the interaction. New expressions for persistent currents and Berry's phases are obtained and nontrivial connections between them are derived. Such properties (potentially useful in the area of topological quantum computation) disappear in case of identical particles, something that demonstrates the importance of dealing with a mixture of distinct particles (for possible applications of the derived universal properties). We should add that the GFs determined here could be useful for a systematic development of solutions for arbitrary interaction (through iteration and the development of the wavefunctions in a Born series). Even for a delta-function interaction, however, the analytical forms for the wavefunctions could be used as input for effective two-particle theories in multiple-connected spaces (both for normal and superconducting rings).

The present paper makes extensive reference to [17], omitting the tedious mathematical details (especially on the GF procedure under the new boundary conditions), and presents only results that generalize the previous work. Section 2 briefly describes the system, its separation in CM and relative variables, and then proceeds to formulate the relative problem in terms of an integral equation. Section 3 outlines how the appropriate GFs are determined and presents the resulting exact condition that graphically determines the entire energy spectrum of the two-particle problem. Section 4 takes up the issue of probability and electric persistent currents – quantities that are studied with respect to combined variation of the several parameters characterizing the system, and two types of adiabatic cyclic processes are identified, with the associated Berry's phases being determined and shown to be directly related to the persistent currents. The results in this Section are universal – independent of the form of the interaction. Section 5 presents our conclusions.

## 2. The system and the method

We consider two spinless particles of arbitrary masses $m_1$ and $m_2$ with charges $q_1$ and $q_2$, respectively, moving along a circular 1D ring of radius $R$, threaded by a magnetic flux $\Phi$. The position of each particle is described by angular variables $\varphi_1$ and $\varphi_2$ (see figure 1 of Ref. [17]).

The two particles interact through an interaction $U(\varphi_1 - \varphi_2)$ which as in Ref. [17] is taken as a contact interaction of the form $U \delta(\varphi_1 - \varphi_2)$ with $U$ a real constant (of either sign) with dimensions of energy. (Contacts are actually periodic in the relative variable $(\varphi_1 - \varphi_2)$ with period $2\pi$, and this determines, through Bloch's theorem, the correct boundary conditions to be imposed at the ends of interval $(-\pi, \pi]$, see next Section).

The Schrödinger equation describing all the eigenfunctions $\psi(\varphi_1, \varphi_2)$ of the system, with a gauge choice of a tangential vector potential $\mathbf{A}$ with uniform magnitude $A = \Phi/2\pi R$, if written in terms of CM ($\Phi_c$) and relative ($\varphi$) variables, defined by

$$\varphi = \varphi_1 - \varphi_2 \qquad \Phi_c = \frac{m_1 \varphi_1 + m_2 \varphi_2}{(m_1 + m_2)}, \qquad (2.1)$$

and if $\psi(\varphi_1, \varphi_2)$ is represented as a product of $\psi(\varphi_1, \varphi_2) = \Psi_c(\Phi_c)\Phi(\varphi)$, is transformed to the form

$$\left[\frac{1}{2M_c}\left(-i\hbar\frac{\partial}{R\partial\Phi_c} - \frac{Q_c}{c}A\right)^2 + \frac{1}{2\mu_r}\left(-i\hbar\frac{\partial}{R\partial\varphi} - \frac{Q_r}{c}A\right)^2 + U\delta(\varphi)\right]\Psi_c(\Phi_c)\Phi(\varphi) \\ = E\psi_c(\Phi_c)\Phi(\varphi), \qquad (2.2)$$

where the total energy of the system $E$ is $E = E_r + E_c$, with $E_r$ the energy of the relative (internal) subsystem, $E_c$ the energy of the CM motion, and

$$M_c = m_1 + m_2, \quad \mu_r = \frac{m_1 m_2}{m_1 + m_2}, \quad Q_c = q_1 + q_2, \quad Q_r = \frac{q_1 m_2 - q_2 m_1}{m_1 + m_2}. \qquad (2.3)$$

(Although the first three quantities above are of a well-known meaning and significance, we should also note at the outset the interesting quantity $Q_r$ : (i) in case $m_1 = m_2$ and $q_1 = q_2 = q$, $Q_r$ is equal to $q$, (ii) in case $m_2 \gg m_1$, $Q_r$ is equal to the charge of the lighter particle, and (iii) if the condition $q_1/m_1 = q_2/m_2$ holds, $Q_r = 0$, and then we note that the AB influence of the enclosed flux on the relative problem disappears (a semiclassical interpretation of this being given later)).

First, the CM problem corresponds to a single free particle on the ring with mass $M_c$ and charge $Q_c$, with the allowed energies and eigenfunctions for the CM motion being

$$E_c = \frac{\hbar^2}{2M_c R^2}\left(N - \frac{Q_c}{hc}\Phi\right)^2 \quad N = 0, \pm 1, \pm 2, \ldots \qquad \Psi_c(\Phi_c) = C_1 \exp(iN\Phi_c) = \frac{1}{\sqrt{2\pi}}\exp(iN\Phi_c). \quad (2.4)$$

The equation for the relative motion can then be transformed to the form

$$\left(-i\frac{\partial}{\partial\varphi} - f\right)^2 \Phi(\varphi) - B\Phi(\varphi) = -\frac{U}{\Delta}\delta(\varphi)\Phi(\varphi), \qquad (2.5)$$

where a dimensionless internal energy $B$ and a dimensionless internal flux $f$ are defined through

$$B = \frac{E_r}{\Delta} = \frac{E - E_c}{\Delta}, \quad f = \frac{Q_r \Phi}{hc}, \quad \Delta = \frac{\hbar^2}{2\mu_r R^2}. \qquad (2.6)$$

As in Ref. [17], we can find the solution of (2.5) using the GF procedure, namely

$$\Phi(\varphi) = \int_{-\pi}^{\pi} d\varphi' G(\varphi,\varphi') \frac{U}{\Delta} \delta(\varphi') \Phi(\varphi') \qquad (2.7)$$

where an appropriate GF for this problem should satisfy Eq. (2.10) of Ref. [17], but with appropriate boundary conditions (to be determined below).

### 3. The Green's function procedure and the energy spectrum

To find the proper boundary conditions for $G(\varphi,\varphi')$ at the ends of the interval $(-\pi,\pi]$, it is useful, as in [17], to temporarily extend the definition of $\varphi$ to the region $-\infty < \varphi < \infty$, where the potential can be represented in the form $V(\varphi) = U \sum_{n=-\infty}^{\infty} \delta(\varphi - 2\pi n)$. Following the standard approach [13, 14] we set

$$\Phi(\varphi) = \exp(if\varphi)\chi(\varphi), \qquad (3.1)$$

so that $f$ is "gauged away" from (2.5) and $\chi(\varphi)$ satisfies a simple Schrödinger equation in a periodic potential with period $2\pi$, having therefore solutions with a Bloch form [13, 17], namely

$$\chi(\varphi) = \exp(ik\varphi)u(\varphi), \qquad (3.2)$$

where $u(\varphi)$ is $2\pi$ – periodic, and $k$ is a real dimensionless number (with $\hbar k$ being the analog of "crystal momentum") that lies within the first Brillouin zone

$$-\frac{1}{2} < k \leq \frac{1}{2}. \qquad (3.3)$$

Using Eq. (2.1), Eq. (2.4), Eq. (3.1) and Eq. (3.2), the total wavefunction can be written in the form

$$\psi(\varphi_1, \varphi_2) = \psi(\varphi, \Phi_c) = C\, e^{iN\frac{(m_1\varphi_1 + m_2\varphi_2)}{(m_1+m_2)}} e^{i(f+k)(\varphi_1 - \varphi_2)} u(\varphi_1 - \varphi_2). \qquad (3.4)$$

(3.4) must be single-valued with respect to $\varphi_1$ and $\varphi_2$ separately, and this leads to the relations

$$\frac{m_1 N}{m_1 + m_2} + f + k = n_1, \qquad \frac{m_2 N}{m_1 + m_2} - f - k = n_2, \qquad (3.5)$$

where $n_1, n_2$ are arbitrary integers. It follows from Eq. (3.5) that $N = n_1 + n_2$ is an integer (with $\hbar N$ being the total angular momentum $L_z$ of the pair) and that $f + k = \frac{m_2}{m_1+m_2}n_1 - \frac{m_1}{m_1+m_2}n_2 = l$. (In case of equal masses $m_1 = m_2$, $l$ turns out to be a half-integer $n/2$, as in Ref. [17]).

The relative wavefunction can therefore be written in the form $\Phi(\varphi) = \exp(il\varphi)u(\varphi)$ with $u(\varphi)$ being $2\pi$ – periodic. The boundary conditions that we need to impose on $\Phi(\varphi)$ are therefore

$$\Phi(\pi) = \exp(i2l\pi)\Phi(-\pi); \qquad \Phi'(\pi) = \exp(i2l\pi)\Phi'(-\pi), \qquad (3.6)$$

and these are the same type of boundary conditions that the GF is also required to satisfy.

By then following the standard mathematical (matching) procedure we obtain an analytical expression for $G(\varphi,\varphi')$ (see Ref. [17]) and then with use of (2.7) we determine all the relative wavefunctions in closed form. By then imposing a self-consistency condition (in the limit $\varphi \to 0$) as in Ref. [17], we obtain after some trigonometric manipulations the following transcendental equation for the entire energy spectrum of the system

$$\frac{\operatorname{Sin}(2\pi\sqrt{B})}{\operatorname{Cos}(2\pi(f-l)) - \operatorname{Cos}(2\pi\sqrt{B})} = \frac{2\Delta\sqrt{B}}{U}; \quad l = n_1 - \frac{m_1}{m_1+m_2}N = f + k, \quad n_1 = 0, \pm 1, \pm 2, \ldots \qquad (3.7)$$

This is a general condition that determines the whole set of allowed energies of this two-particle problem (in combination with the definition of $B$ (see (2.6)), and with $E_c$ given by (2.4)). For transforming it to a more practical form, it is useful to introduce new dimensionless variables

$$\mu = \frac{m_1}{m_2}, \quad \lambda = \frac{q_1}{q_2}, \quad \Delta_0 = \frac{\hbar^2}{m_2 R^2}, \qquad (3.8)$$

and the energy spectrum equation (3.8) can then be written in the final form (with $\Phi_0 = hc/|e|$)

$$\frac{\mathrm{Sin}\left(2\pi\sqrt{\frac{2\mu}{1+\mu}\frac{E}{\Delta_0} - \frac{\mu}{(1+\mu)^2}\left(N - (\lambda+1)\frac{q_2}{|e|}\frac{\Phi}{\Phi_0}\right)^2}\right)}{\mathrm{Cos}\left(2\pi\left(\frac{\lambda-\mu}{1+\mu}\frac{q_2}{|e|}\frac{\Phi}{\Phi_0} + \frac{\mu}{1+\mu}N\right)\right) - \mathrm{Cos}\left(2\pi\sqrt{\frac{2\mu}{1+\mu}\frac{E}{\Delta_0} - \frac{\mu}{(1+\mu)^2}\left(N - (\lambda+1)\frac{q_2}{|e|}\frac{\Phi}{\Phi_0}\right)^2}\right)} \qquad (3.9)$$

$$= \frac{1+\mu}{\mu}\frac{\Delta_0}{U}\sqrt{\frac{2\mu}{1+\mu}\frac{E}{\Delta_0} - \frac{\mu}{(1+\mu)^2}\left(N - (\lambda+1)\frac{q_2}{|e|}\frac{\Phi}{\Phi_0}\right)^2}.$$

We have determined the energies graphically (by plotting the left-hand-side and the right-hand-side of (3.9) as functions of $E$ in a common system of axes, and locating the points of their intersections).

The results for the energy spectrum of the system for different values of various parameters are presented in figure 1 (where we have taken $m_2 = m_0$, with $m_0$ being the bare electron mass). As shown in figure 1a, for every value of $f$ there is a critical value of interaction potential $U_c$; for $U < U_c$ the ground state energy becomes negative. From the equality of the derivatives of both sides of Eq. (3.7) at $B = 0$ the form of the critical value $U_c$ for binding turns out to be

$$U_c = -\frac{\Delta}{\pi}\left(1 - \mathrm{Cos}\left[2\pi\left(f + \frac{m_1}{m_1+m_2}N\right)\right]\right). \qquad (3.10)$$

The existence of a critical interaction is a result of the competition between the delocalizing influence of the enclosed magnetic flux and the binding influence of the attractive interaction between the particles. Eq. (3.9) gives the opportunity to also study excited states and optical properties as well.

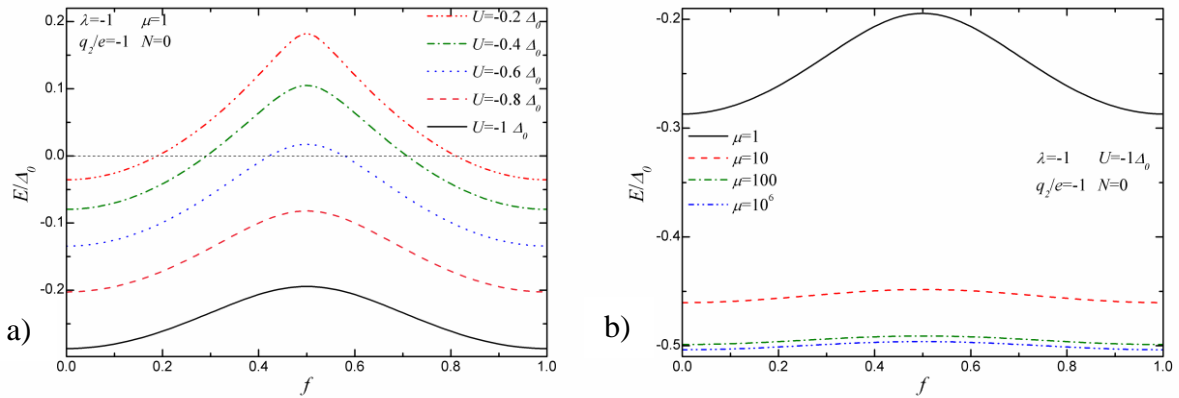

**Figure 1.** (Color online) The dependence of the ground state energy of the system on flux parameter $f$: a) for different (negative) values of parameter $U$; b) for different values of $\mu$.

## 4. Persistent currents and Berry's phases

Apart from optical transitions associated with the above energy spectra, we can also determine other physical quantities that can also be directly measured in an experiment. Let us focus here on probability and electric densities and currents for our model system. Because the number of particles is more than one, these (local) quantities are given as expectation values of appropriate many-body operators, defined by Eqs (6.1)-(6.4) of Ref. [17]. Here we just provide the electric current operator

$$\hat{J}_{el}(\bar{\varphi}) = \frac{1}{2R}\sum_{i=1}^{N} q_i \left( \frac{\hat{p}_i}{m_i} \delta(\bar{\varphi}-\varphi_i) + \delta(\bar{\varphi}-\varphi_i)\frac{\hat{p}_i}{m_i} \right) - \frac{A}{Rc}\frac{q_i^2}{m_i}\delta(\bar{\varphi}-\varphi_i), \quad (4.1)$$

to remind the reader that $\bar{\varphi}$ is the only variable to survive at the end, describing the point on the ring where each quantity is locally determined (the $\varphi_i$ being merely integration variables in the expectation values to be taken (see below)). By taking $N = 2$ and by using the analytical expressions found for the total wavefunctions of our system, we obtain by direct evaluation of the expectation values the following results (all quantities turning out to be spatially uniform, as expected for eigenstates):

$$\rho(\bar{\varphi}) = \langle \hat{\rho}(\bar{\varphi}) \rangle = \frac{2}{2\pi R}, \quad (4.2)$$

$$\rho_{el}(\bar{\varphi}) = \langle \hat{\rho}_{el}(\bar{\varphi}) \rangle = \frac{q_1 + q_2}{2\pi R} = \frac{Q_c}{2\pi R}, \quad (4.3)$$

$$J(\bar{\varphi}) = \langle \hat{J}(\bar{\varphi}) \rangle = \frac{1}{2\pi R^2}\frac{2}{M_c}\left\{ \hbar N - \frac{Q_c}{2\pi c}\Phi \right\} - \frac{2}{R^2 M_c \mu_r}\left( \pm\sqrt{M_c^2 - 4M_c\mu_r} \right)$$

$$\times \left\{ \frac{B\hbar Sin(2\pi(l-f))Sin(2\pi\sqrt{B})}{-4\pi\sqrt{B} + 2Cos(2\pi(l-f))\left(2\pi\sqrt{B}Cos(2\pi\sqrt{B}) - Sin(2\pi\sqrt{B})\right) + Sin(4\pi\sqrt{B})} \right\},$$

(4.4)

$$J_{el}(\bar{\varphi}) = \langle \hat{J}_{el}(\bar{\varphi}) \rangle = \frac{1}{2\pi R^2}\frac{Q_c}{M_c}\left\{ \hbar N - \frac{Q_c}{2\pi c}\Phi \right\} - \frac{2Q_r}{R^2 \mu_r}$$

$$\times \left\{ \frac{B\hbar Sin(2\pi(l-f))Sin(2\pi\sqrt{B})}{-4\pi\sqrt{B} + 2Cos(2\pi(l-f))\left(2\pi\sqrt{B}Cos(2\pi\sqrt{B}) - Sin(2\pi\sqrt{B})\right) + Sin(4\pi\sqrt{B})} \right\},$$

(4.5)

where $+\sqrt{M_c^2 - 4M_c\mu_r}$ corresponds to the case $m_2 \geq m_1$ and $-\sqrt{M_c^2 - 4M_c\mu_r}$ to the case $m_2 \leq m_1$. The above results are very general, but it is useful to discuss some special cases: first, by taking $m_1 = m_2 = m$ and $q_1 = -q_2 = q$, we will have $M_c^2 - 4M_c\mu_r = 0$ and $Q_c = 0$, so that the persistent currents for such an exciton turn out to be given by

$$I_{prob}^{pers} = \frac{\hbar N}{2\pi m R^2}, \quad (4.6)$$

$$I_{elect}^{pers} = \frac{4q}{R^2 m}\left\{ \frac{B\hbar Sin(2\pi(l-f))Sin(2\pi\sqrt{B})}{-4\pi\sqrt{B} + 2Cos(2\pi(l-f))\left(2\pi\sqrt{B}Cos(2\pi\sqrt{B}) - Sin(2\pi\sqrt{B})\right) + Sin(4\pi\sqrt{B})} \right\}, \quad (4.7)$$

which after some tedious manipulations can be shown to be in agreement with Eq. (6.28) of Ref. [17]. Another interesting case to consider is the case when $q_1/m_1 = q_2/m_2$, so that $Q_r = 0$. In that case the AB effect disappears from the internal problem, while it is still operating at the level of the CM motion, and as a consequence $I_{pers}^{elect}$ is described by only the first CM-term of (4.5). This was actually expected by (2.2) as mentioned earlier, a semiclassical interpretation being that, when the above condition holds between masses and charges, then it is expected that we have the formation of a mini-crystal of 2 particles, antidiametrically placed in the circle: the internal Physics of this pair does not change (semiclassically) with $f$ (for, if i.e. we vary $f$ with time, then we have a force acting on each of the two charges due to the Electromagnetic Induction ($F_i = q_i E_{ind}$), and (when the above condition holds) the associated accelerations are equal in magnitude ($q_1 E_{ind}/m_1 = q_2 E_{ind}/m_2$), with the result that the internal relative motion is not changed by the variation of $f$).

Figures 2 and 3 show the dependence of the electric persistent current on various different parameters. In figure 3 the oscillating behavior should be noted, not present in the case of non-interacting particles.

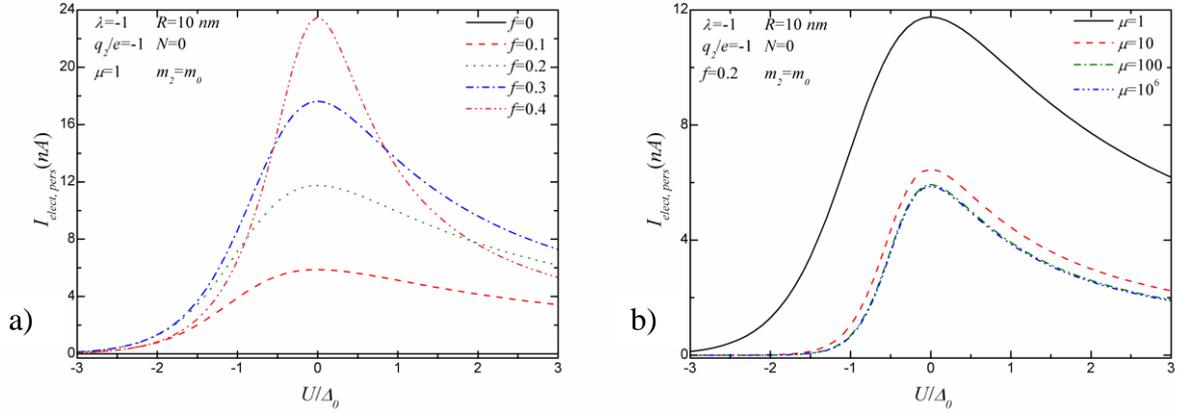

**Figure 2.** (Color online) The dependence of the electric persistent current on interaction potential $U$ for different values of: a) $f$; b) $\mu$.

Let us also provide some more general results for persistent currents (independent of the form of interaction) that will be used later (see (4.15) and (4.16)). By taking general expectation values of (4.1) and integrating some of the terms by parts we obtain

$$I_{pers}^{prob} = \frac{1}{2\pi m_1 R}\left\langle \hat{p}_1 - \frac{q_1}{c} A(\varphi_1,t)\right\rangle + \frac{1}{2\pi m_2 R}\left\langle \hat{p}_2 - \frac{q_2}{c} A(\varphi_2,t)\right\rangle, \quad (4.8)$$

$$I_{pers}^{elect} = \frac{q_1}{2\pi m_1 R}\left\langle \hat{p}_1 - \frac{q_1}{c} A(\varphi_1,t)\right\rangle + \frac{q_2}{2\pi m_2 R}\left\langle \hat{p}_2 - \frac{q_2}{c} A(\varphi_2,t)\right\rangle, \quad (4.9)$$

which if written in terms of CM and relative variables give the final forms

$$I_{pers}^{prob} = \frac{2}{2\pi M_c R}\left\langle \hat{p}_{\Phi_c} - \frac{Q_c}{c} A(t)\right\rangle + \frac{1}{2\pi R M_c \mu_r}\left(\pm\sqrt{M_c^2 - 4M_c\mu_r}\right)\left\langle \hat{p}_\varphi - \frac{Q_r}{c} A(t)\right\rangle, \quad (4.10)$$

$$I_{pers}^{elect} = \frac{Q_c}{2\pi M_c R}\left\langle \hat{p}_{\Phi_c} - \frac{Q_c}{c} A(t)\right\rangle + \frac{Q_r}{2\pi R \mu_r}\left\langle \hat{p}_\varphi - \frac{Q_r}{c} A(t)\right\rangle. \quad (4.11)$$

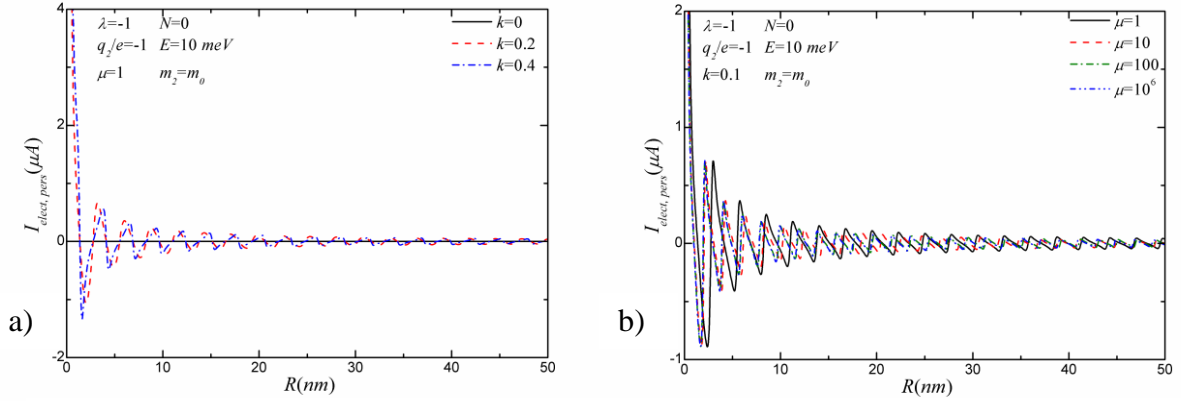

**Figure 3.** (Color online) The dependence of the electric persistent current on ring radius for different values of: a) $k$ ; b) $\mu$.

Finally, let us introduce certain cyclic adiabatic processes for which the corresponding Berry's phases can be determined in closed form, and they are related to the above currents. In so doing, we will find relations that are actually more general than the above case of a delta-interaction. As is well known a Berry's phase [18] for a cyclic adiabatic process is determined by

$$\gamma(c) = i\oint_C \langle \Psi_n(\mathbf{R_0}) | \nabla_{\mathbf{R_0}} \Psi_n(\mathbf{R_0}) \rangle \cdot d\mathbf{R_0} , \qquad (4.12)$$

where $\mathbf{R_0}$ is a t-dependent parameter that appears in the Hamiltonian of the system and is varied adiabatically in a cyclic path $C$ in parameter space. (Below we will introduce such parameters in real space that make such cyclic paths, by exploiting essentially the ring-topology of our system). The idea on which the following calculation is based is actually a generalization of a largely unnoticed paper by Berry [19] (that referred to a single particle), where a geometric phase due to a slow overall rotation of the ring turns out to consist of 2 terms: the 1st is the usual AB phase (the one that would result from adiabatic transport of a box with rigid walls [18]), and the 2nd is directly related to the electric current. It is worth emphasizing here on the possible use of a geometric phase as a "theoretical detector" of conduction (in an interacting system, in the sense of [20]), and this should be contrasted to the theory of Macroscopic Polarization [21, 22] where generalized Berry's phases also appear (but with adiabatic parameters defined in $k$- rather than in real space).

Because our Hamiltonian consists of two additive terms (with one describing the CM and the other the internal motion) and these two terms commute, it turns out that, if we use adiabatic parameters that couple separately to the CM and the relative subsystem, the overall Berry's phase can be written as a sum of two separate parts, namely

$$\gamma(c) = \gamma_c(c) + \gamma_r(c) = i\oint \left\langle \Psi(\Phi_c, R_{01})\Phi(\varphi) \left| \frac{\partial}{\partial R_{01}} \right| \Psi(\Phi_c, R_{01})\Phi(\varphi) \right\rangle dR_{01}$$
$$+ i\oint \left\langle \Psi(\Phi_c)\Phi(\varphi, R_{02}) \left| \frac{\partial}{\partial R_{02}} \right| \Psi(\Phi_c)\Phi(\varphi, R_{02}) \right\rangle dR_{02} \qquad (4.13)$$

where the parameter $R_{01}$ has been assumed to be coupled to the CM only, and $R_{02}$ to be coupled only to the relative subsystem. First, as an $R_{01}$, we introduce a parameter $\varphi_0(t)$ that couples only to the CM motion, namely, we just replace $\Phi_c$ with $\Phi_c - \varphi_0(t)$ (and leave $\varphi$ unchanged), so that the corresponding Berry's phase (for an adiabatic change of $\varphi_0(t)$ from 0 to $2\pi$) is given by

$$\gamma_c(c) = i\int_0^{2\pi} \left\langle \Psi(\Phi_c - \varphi_0(t), t)\Phi(\varphi, t) \left| \frac{\partial \Psi(\Phi_c - \varphi_0(t), t)}{\partial \varphi_0(t)} \right| \Phi(\varphi, t) \right\rangle d\varphi_0(t). \qquad (4.14)$$

(This actually describes a slow overall rotation (by $2\pi$) of the whole circular ring around its center). By evaluating expressions (4.10) and (4.11) explicitly for this case we obtain

$$I_{prob \atop pers} = \frac{2}{2\pi R M_c}\left(\frac{-i\hbar}{R}\right)\iint \Psi^*(\Phi_c - \varphi_0(t),\varphi,t)\frac{\partial \Psi(\Phi_c - \varphi_0(t),\varphi,t)}{\partial \Phi_c}d\Phi_c d\varphi - \frac{2Q_c}{4\pi^2 R^2 c M_c}\Phi$$
$$-\frac{1}{2\pi R}\frac{\Delta m}{\mu_r M_c}\left(\frac{-i\hbar}{R}\right)\iint \Psi^*(\Phi_c - \varphi_0(t),\varphi,t)\frac{\partial \Psi(\Phi_c - \varphi_0(t),\varphi,t)}{\partial \varphi}d\Phi_c d\varphi + \frac{1}{4\pi^2 R^2 c}\frac{\Delta m Q_r}{\mu_r M_c}\Phi, \quad (4.15)$$

$$I_{elect \atop pers} = \frac{Q_c}{2\pi R M_c}\left(\frac{-i\hbar}{R}\right)\iint \Psi^*(\Phi_c - \varphi_0(t),\varphi,t)\frac{\partial \Psi(\Phi_c - \varphi_0(t),\varphi,t)}{\partial \Phi_c}d\Phi_c d\varphi - \frac{Q_c^2}{4\pi^2 R^2 c M_c}\Phi$$
$$+\frac{1}{2\pi R}\frac{Q_r}{\mu_r}\left(\frac{-i\hbar}{R}\right)\iint \Psi^*(\Phi_c - \varphi_0(t),\varphi,t)\frac{\partial \Psi(\Phi_c - \varphi_0(t),\varphi,t)}{\partial \varphi}d\Phi_c d\varphi - \frac{1}{4\pi^2 R^2 c}\frac{Q_r^2}{\mu_r}\Phi, \quad (4.16)$$

where $\Delta m = m_1 - m_2$. By then using the obvious equality

$$\frac{\partial \Psi(\Phi_c - \varphi_0(t),\varphi,t)}{\partial \Phi_c} = -\frac{\partial \Psi(\Phi_c - \varphi_0(t),\varphi,t)}{\partial \varphi_0(t)}, \quad (4.17)$$

we see that the corresponding Berry's phase takes the form

$$\gamma_c(c) = \frac{2\pi R^2}{\hbar}\frac{\Delta m}{\Delta q}\left\{\int_0^{2\pi} I_{elect \atop pers} d\varphi_0(t)\right\} + \frac{2\pi R^2}{\hbar}\frac{Q_r M_c}{\Delta q}\left\{\int_0^{2\pi} I_{prob \atop pers} d\varphi_0(t)\right\} + \frac{Q_c}{\hbar c}\Phi, \quad (4.18)$$

where $\Delta q = q_1 - q_2$ and we have assumed that $q_1 \neq q_2$. In case $q_1 = q_2 = q$ and $m_1 = m_2 = m$, a more careful calculation finally gives

$$\gamma_c(c) = \frac{2\pi R^2 m}{\hbar}\left\{\int_0^{2\pi} I_{prob \atop pers} d\varphi_0(t)\right\} + \frac{2q}{\hbar c}\Phi = \frac{2\pi R^2}{\hbar}\frac{m}{q}\left\{\int_0^{2\pi} I_{elect \atop pers} d\varphi_0(t)\right\} + \frac{2q}{\hbar c}\Phi \quad (4.19)$$

(and in this case of $q_1 = q_2 = q$ it is easy to see from (4.15) and (4.16) that $I_{elect \atop pers} = q I_{prob \atop pers}$).

By following a similar procedure for the relative motion, namely by introducing (as an $R_{02}$) a parameter $\varphi_0(t)$ now coupled only to the relative problem (as in Ref. [17]), namely by just replacing $\varphi$ with $\varphi - \varphi_0(t)$ (and by leaving $\Phi_c$ unchanged), and by following similar steps (for an adiabatic change of $\varphi_0(t)$ from 0 to $2\pi$), we finally obtain the Berry's phase for the internal motion

$$\gamma_r(c) = \frac{4\pi R^2}{\hbar}\frac{\mu_r}{\Delta q}\left\{\int_0^{2\pi} I_{elect \atop pers} d\varphi_0(t)\right\} - \frac{2\pi R^2}{\hbar}\frac{\mu_r Q_c}{\Delta q}\left\{\int_0^{2\pi} I_{prob \atop pers} d\varphi_0(t)\right\} + 2\pi f \quad (4.20)$$

where again we assumed $q_1 \neq q_2$. For the case $q_1 = q_2 = q$ and $m_1 = m_2 = m$, it is interesting to note that the entire above scheme collapses: no connection exists whatsoever between the above relative-Berry's phase and any of the currents. The character of the charged mixture seems to be crucial in order to derive the above results, a rather important point to stress for possible applications.

For the special case of an exciton ($q_1 = -q_2 = |e|$) and $m_1 = m_2 = m_e$ we obtain

$$\gamma_c(c) = \frac{2\pi R^2 m_e}{\hbar}\left\{\int_0^{2\pi} I_{prob \atop pers} d\varphi_0(t)\right\}, \quad (4.21)$$

$$\gamma_r(c) = \frac{\pi R^2}{\hbar}\frac{m_e}{|e|}\left\{\int_0^{2\pi} I_{elect \atop pers} d\varphi_0(t)\right\} + \frac{|e|}{\hbar c}\Phi, \quad (4.22)$$

showing that now the CM-Berry's phase is only related to the probability current, while the relative-Berry's phase has one AB term, and one term related to the electric current only.

It should be reminded here that expressions (4.18)-(4.22) have been derived independently of the form of interaction; they therefore represent universal patterns of behavior for this system, and all the above results describe the connection between Berry's phases and persistent currents *for any kind of interaction potential*. This is one of the central results of this paper, since Berry's phases are seen to be directly related to measurable quantities (e.g. electric current) that can be easily monitored, something that could possibly be useful in the area of topological quantum computation.

**5. Conclusions**
In the present work the problem of a system of two interacting (spinless) particles with arbitrary masses and charges moving in a doubly-connected space and in the presence of an AB flux was exactly solved using a GF procedure. The model problem consisted initially of a system of two interacting charged particles moving in a 1D ring, threaded by a magnetic flux, and with a contact interaction, although at the end, several results were given that were found to be independent of the form of the interaction.

The energy spectrum and the associated eigenstates together with other physical properties of the system were determined in closed analytical forms, and could be useful in studies of transport and optical properties of interacting particles in multiple-connected spaces, such as nanorings.

Two kinds of Berry's phases associated with CM and relative motions were introduced and investigated, and were shown to be directly linked to the electric and probability (persistent) currents, in patterns independent of the form of the interaction.

The obtained Berry's phases are directly related to measurable quantities that can be easily manipulated in an experiment, and this could prove useful in the area of quantum computation, where recently geometric phases are called to provide topological stability against decoherence.